\documentclass[prl,aps,twocolumn,floatfix,showpacs]{revtex4}
\usepackage{graphicx,psfrag,amsmath,calc,amssymb}
\usepackage{color}

\begin{document}
\title{Superfluid and Fermi liquid phases of Bose-Fermi mixtures
in optical lattices}
\author{Kaushik Mitra, C. J. Williams and  C. A. R. S{\'a} de Melo}
\affiliation{Joint Quantum Institute, University of Maryland, College Park, Maryland 20742,\\ 
and National Institute of Standards and Technology, Gaithersburg, Maryland 20899}
\date{\today}

\begin{abstract}
We describe interacting mixtures of ultracold bosonic and fermionic atoms in harmonically 
confined optical lattices. For a suitable choice of parameters we study the emergence of
superfluid and Fermi liquid (non-insulating) regions out of Bose-Mott and Fermi-band insulators,
due to finite Boson and Fermion hopping. We obtain the shell structure for the
system and show that angular momentum can be transferred to the non-insulating regions 
from Laguerre-Gaussian beams, which combined with Bragg spectroscopy can reveal all 
superfluid and Fermi liquid shells.  
\end{abstract}

\maketitle

%
%
Fermi and Bose degenerate quantum gases and liquids are amazing systems,
which have revealed individually several macroscopic quantum phenomena.
For instance, superfluidity is known to exist in neutral liquids such 
as $^4$He (boson) and in $^3$He (fermion), as well as in a variety 
of electronic materials studied in standard condensed matter physics. 
The role of quantum statistics and interactions is of fundamental importance 
to understand the phases emerging from purely bosonic
or purely fermionic systems, and a substantial amount of understanding of these
individual Bose or Fermi systems can be found in the atomic 
and condensed matter  
physics literature.
However, new frontiers can be explored when mixtures of bosons and fermions are 
produced in harmonic traps or optical lattices. An important example of the richness 
of quantum degenerate Bose-Fermi mixtures was revealed in standard condensed matter 
systems,
where for fixed $^4$He density and increasing 
amounts of $^3$He, the critical temperature for superfluidity is reduced, 
and below a tricritical point phase separation appears~\cite{reppy-1967}. 
In this Bose-Fermi mixture of standard condensed matter physics, essentially 
the only control parameter is the ratio between the densities of $^4$He and $^3$He.

In atomic physics, a spectacular degree of control has been achieved in Bose-Fermi
mixtures, where not only the ratio between densities of bosons and fermions 
can be adjusted, but also the interactions between fermions and bosons 
can be controlled through the use of Feshbach resonances~\cite{bongs-2006},
as demonstrated in mixtures of harmonically trapped Bose-Fermi polarized Fermion mixtures of
$^{40}$K and $^{87}$Rb. Furthermore, these same atoms have been succesfully 
loaded into optical lattices~\cite{esslinger-2006} and have produced a system
that has no counterpart in standard condensed matter systems. By controlling
the depths of optical lattices we can change not only the interactions between boson
and fermions, but also their hopping from site to site, thus allowing the exploration of
a very rich phase space, where supersolid and phase separated states have been 
suggested~\cite{blatter-2003}.
The list of Bose-Fermi mixtures in atomic physics is growing, and 
include systems where the masses are close like $^6$Li and $^7$Li, $^{39}$K and $^{40}$K,
or $^{172}$Yb and $^{173}$Yb; or systems where the masses are quite different like
$^{6}$Li and $^{40}$K, $^{7}$Li and $^{39}$K, $^6$Li and $^{23}$Na, $^6$Li and $^{87}$Rb,
$^{40}$K and $^{23}$Na, or $^{40}$K and $^{87}$Rb.
This suggests a wide possibility of regimes that can be reached by tuning interactions, 
density and geometry, which is not is possible in ordinary condensed matter physics.

A few studies of quantum phases of Bose-Fermi mixtures have focused on homogeneous three
dimensional systems with~\cite{lewenstein-2004, batrouni-2007}, 
and without optical lattices~\cite{viverit-2000}.
However, only one effort focused on harmonically confined 
optical lattices~\cite{illuminati-2004}. Most of the descriptions of
Bose-Fermi mixtures in optical lattices have relied on numerical methods using
either Gutzwiller projection~\cite{lewenstein-2004, illuminati-2004}
or quantum Monte Carlo~\cite{batrouni-2007} techniques. In this paper, we present 
a fully analytical theory of boson and spin polarized fermion mixtures in harmonically confined optical lattices
by using degenerate perturbation theory for finite hopping in conjunction with 
the local density approximation. This work provides insight 
into the phase diagram of Bose-Fermi mixtures, and into the detection of superfluid
and Fermi liquid shells at low temperatures.

This paper we analyses in detail the regime where the hopping parameters 
of bosons and fermions are comparable and the repulsion between bosons 
and fermions is a substantial fraction
of the boson-boson repulsion. In this case, the system presents
regions of (I) coexisting Bose-Mott and Fermi-band insulator, 
(II) coexisting Bose-Mott insulator and Fermi liquid, 
(III) Bose-Mott insulator,  and  
(IV) Bose superfluid, as shown in Fig. 1.
We compute analytically the boundaries between various phases, and
obtain the spatially dependent boson and fermions filling fractions 
in each region. Although one can envisage other situations where, for example, one can have 
coexistence of superfluid and Fermi band insulator, we confine our discussion to the situation  
above for the sake of simplicity.   
Finally, we propose a detection method of the shell strucuture of Bose-Fermi 
mixtures by using Laguerre-Gaussian beams~\cite{helmerson-2007} 
and Bragg spectroscopy~\cite{raman-2006}, 
where angular momentum is transferred only to regions with extended states such
as the superfluid and Fermi-liquid shells.

To describe Bose-Fermi mixtures in harmonically confined
square (2D) or cubic (3D) optical lattices we start with the Hamiltonian
\begin{eqnarray}
\hat{H}&=&\label{eq:bfh}
K_B + K_F -\sum_{\bf r} \mu_{B} ({\bf r}) {\hat n}_{B} ({\bf r}) 
-\sum_{\bf r} \mu_F ({\bf r}) {\hat n}_{F} ({\bf r}) \nonumber \\
&+& \frac{U_{BB}}{2} 
\sum_{\bf r}  {\hat n}_{B} ({\bf r}) \left[ {\hat n}_{B} ({\bf r}) - 1  \right]
+ U_{BF}\sum_{\bf r} {\hat n}_{B} ({\bf r})  {\hat n}_{F} ({\bf r}) \nonumber, 
\end{eqnarray}
where 
$K_B = -t_B \sum_{\langle {\bf r}, {\bf r}^{\prime}\rangle }  
b_{\bf r}^\dagger b_{ {\bf r}^{\prime} } $ and
$K_F = -t_F \sum_{\langle {\bf r}, {\bf r}^{\prime} \rangle } 
f_{\bf r}^\dagger f_{ {\bf r}^{\prime} }$
are the kinetic energies of boson and fermions with 
nearest-neighbor hoppings $t_B$ and $t_F$, and
$b^{\dagger}_{\bf r}$ and $f^{\dagger}_{\bf r}$ are the  
bosonic and fermionic creation operators at site ${\bf r}$. 
Here, the lattice sites for bosons and fermions are assumed to be the same,
but the hopping parameters can be different. 
The number operators are 
${\hat n}_{B} ({\bf r}) = b_{\bf r}^\dagger b_{\bf r}$ and
$\hat {n}_{F} ({\bf r}) = f_{\bf r}^\dagger f_{\bf r}$,
and the corresponding local chemical potentials 
are $\mu_F ({\bf r}) = \mu_F - V_F ({\bf r})$
and $\mu_B ({\bf r}) = \mu_B - V_B ({\bf r})$, 
where $V_F ({\bf r})= \Omega_F (r/a)^2/2$ and $V_B ({\bf r}) = \Omega_B (r/a)^2/2$ 
are the harmonically confining potentials and 
$\mu_F$ and $\mu_B$ are the chemical potentials for fermions and
bosons. The origin of the lattice with spacing $a$ is chosen to be at the minimum of 
the harmonically confining potential.
The terms containing $U_{BB}$ $( U_{BF} )$ represent the boson-boson
(boson-fermion) interaction.

When $t_B = t_F = 0$, 
the Hamiltonian ${\hat H} = {\hat H}_0$ is
a sum of single-site contributions, and the eigenstates are tensor products
of number states with state vectors
$
|\psi\rangle=
|n_{B,0}, n_{B,1}, \cdots \rangle
|n_{F,0}, n_{F,1}, \cdots\rangle
$,
with $n_{B,{\bf r}}= 0,1,2,... $ and $n_{F,{\bf r}} = 0,1$ 
representing the occupation number of bosons and fermions
at site ${\bf r}$, respectively. At site ${\bf r}$ the local energy is 
$E_{n_B,n_F}({\bf r})= U_{BB}
n_B(n_B-1)/2+U_{BF}n_Bn_F-\mu_B({\bf r})n_B -\mu_F({\bf r})n_F$. 
For the ground state wavefunction the number of bosons at site ${\bf r}$ 
is determined by $max(0,\lfloor (U_{BB}+\mu_B({\bf r}))/2U_{BB}\rfloor)$ if
$E_{n_B,0}({\bf r})< E_{n_B,1}({\bf r})$ and 
$max(0,\lfloor (U_{BB}+\mu_B({\bf r})-U_{BF})/2U_{BB}\rfloor)$ 
otherwise.  Similarly, the number of fermions
per site is zero if $E_{n_B,0}({\bf r})<E_{n_B,1}({\bf r})$ and
one otherwise. The symbol $\lfloor\cdot\rfloor$ is the floor function.
In the ground state solution shells $(n_B,n_F)$ with $n_B$ bosons
and $n_F$ fermions are formed by those lattice sites ${\bf r}$ for
which the local energy is the same.  For our harmonic traps these
shells are nearly spherically symmetric.  The boundary between shells
$(n_B,n_F)$ and $(n_B\!+\!1,n_F)$ is determined by 
$E_{n_B,n_F}({\bf r})=E_{n_B\!+\!1,n_F}({\bf r})$, leading to the radius 
$R_{B,n_B,n_F} = a\sqrt{\Omega_{n_B,n_F}/\Omega_B}$, where 
$\Omega_{n_B,n_F} = 2(\mu_B - n_B U_{BB} -n_F U_{BF})$. 
Similarly, the boundary between shells with occupation numbers
$(n_B, 0)$ and $(n_B, 1)$ is determined
by equating the local energies 
$E_{n_B, n_F} ({\bf r})$ and $E_{n_B, n_F +1} ({\bf r})$,
leading to the radius
$R_{F,n_B} = a \sqrt{2( \mu_F - n_BU_{BF})/\Omega_F}$.
We consider the number of particles to be sufficiently large
such that the radii of the boundaries are much larger than $a$.


Next, we begin our discussion of finite hoppings by taking first 
$t_B \ne 0$, with $t_F = 0$. The Bose superfluid region emerges
due to kinetic fluctuations at the boundaries between 
the $(n_B, n_F)$ and $(n_B + 1, n_F)$ shells.
At this boundary the local energy 
$E_{n_B + 1, n_F} ({\bf r})$ is degenerate with 
$E_{n_B, n_F} ({\bf r})$. 
To describe the emergence of superfluid regions, 
we introduce the order parameter for superfluidity
$\psi_{B,j} $ via the transformation  
$b_i^\dagger b_j  \to   \psi_{B,i}^* b_j +
b_i^\dagger \psi_{B,j} - {\psi_{B,i}^* \psi_{B,j} }$, 
and then for analytical convenience 
make the continuum approximation $\psi ({\bf r} + {\bf \it a}) =  \psi ({\bf r}) +
  a_i \partial_i \psi ({\bf r})
+ (1/2) a_i a_j \partial_i \partial_j \psi ({\bf r}) $.

In the limit of $U_{BB} \gg t_B$, we can restrict our Hilbert 
space to the number basis states $\vert n_B, n_F \rangle$ and
$\vert n_B + 1, n_F \rangle$, as any contribution from other basis states to the local
energy is of order $t_B^2/U_{BB}$. The hopping term $t_B$ affects
the energies $E_{n_B +1, n_F} ({\bf r})$ and $E_{n_B, n_F} ({\bf r})$ 
by removing their degeneracy, thus creating finite-width superfluid
regions between shells $(n_B + 1, n_F)$ and $(n_B, n_F)$. 
The effective local Hamiltonian then becomes,
\begin{equation}
\label{eqn:h-matrix}
H_{\mathbf{r}}^{\textrm{eff}}=
\left(\begin{array}{cc}
E_{n_B,n_F} ( {\bf r} ) + \Lambda( {\bf r} ) & -\sqrt{n_B + 1}\Delta ( {\bf r} )\\
-\sqrt{n_B + 1}\Delta^{*} ( {\bf r} ) & E_{n_B + 1,n_F}( {\bf r} ) + \Lambda( {\bf r} )
\end{array}\right),
\end{equation}
where $\Lambda({ \bf r})=\frac{1}{2}(\Delta( {\bf r})\psi^{*}( {\bf r} )+cc)$
and $\Delta( {\bf r})= t_B (z\psi( { \bf r})+ {\it a}^{2}\nabla^{2}\psi( {\bf r}))$.
Here, $z$ is the coordination number which depends on the lattice dimension
$d$. 

The eigenvalues of Eq.~(\ref{eqn:h-matrix}) are given by, 
%
$$
\label{eqn:eigenvalues}
E_{\pm} (\mathbf{r}) 
= E_s (\mathbf{r}) \pm\sqrt{ \left[  E_d (\mathbf{r}) \right]^2
+ (n_B + 1)\left|\Delta(\mathbf{r})\right|^{2}},
$$
%
where $E_s(\mathbf{r}) = \left[ E_{n_B +1,n_F)}({\bf r})
+ E_{n_B, n_F} ( {\bf r})\right]/2 + \Lambda( {\bf r})$
is proportional to the sum of the diagonal terms, and
$E_d(\mathbf{r}) = 
\left[ E_{n_B +1, n_F}( {\bf r}) 
- E_{n_B, n_F}( {\bf r})\right]/2 
= ( n_B U_{BB} + n_F U_{BF} - \mu_B ( {\bf r} )/2
$
is proportional to their difference. 
Notice that $E_{-} ( {\bf r} )$ is the lowest local energy
leading to the total ground state energy 
$ {\bf E} = \frac{1}{L^d}\int d {\bf r} E_{-} ( {\bf r} ).$
%

\begin{figure} [htb]
\centerline{ 
\scalebox{0.35} {  
\includegraphics{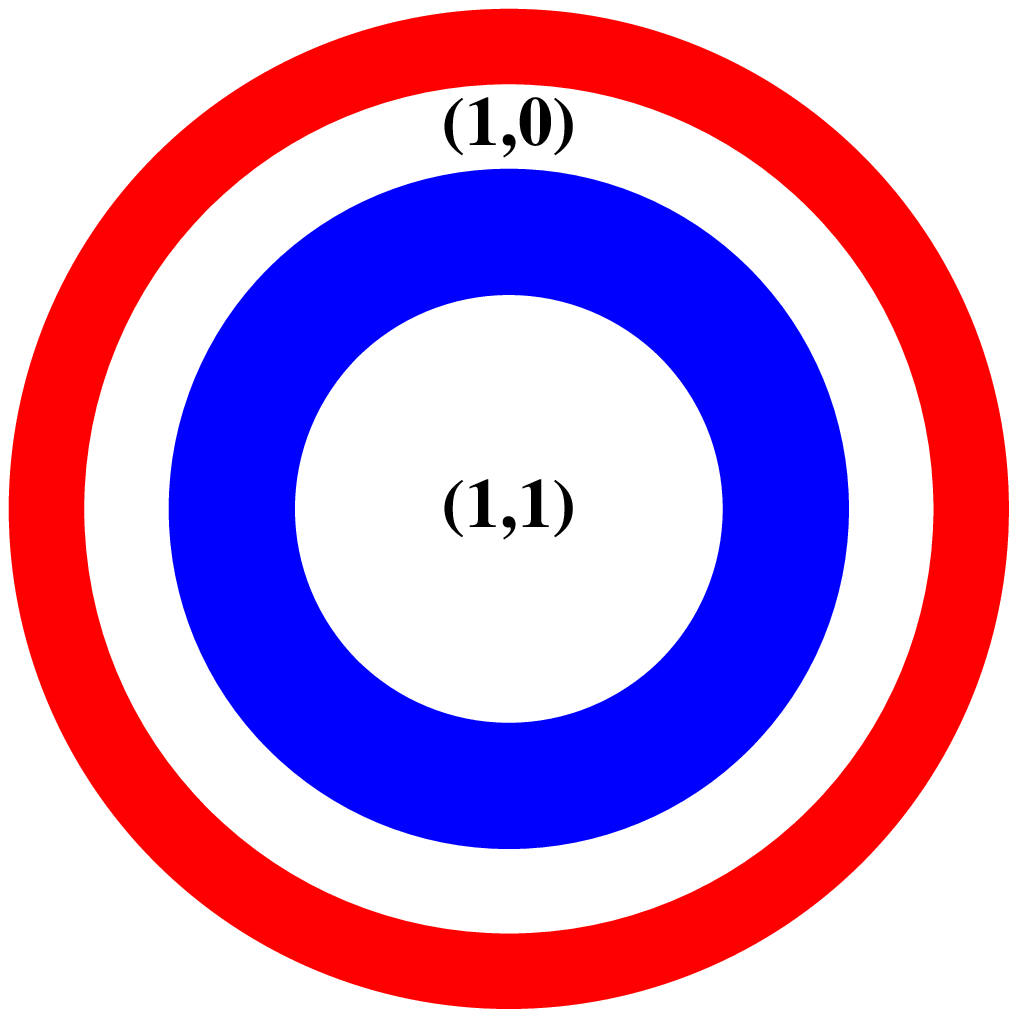}
} 
\scalebox{0.45} {
\includegraphics{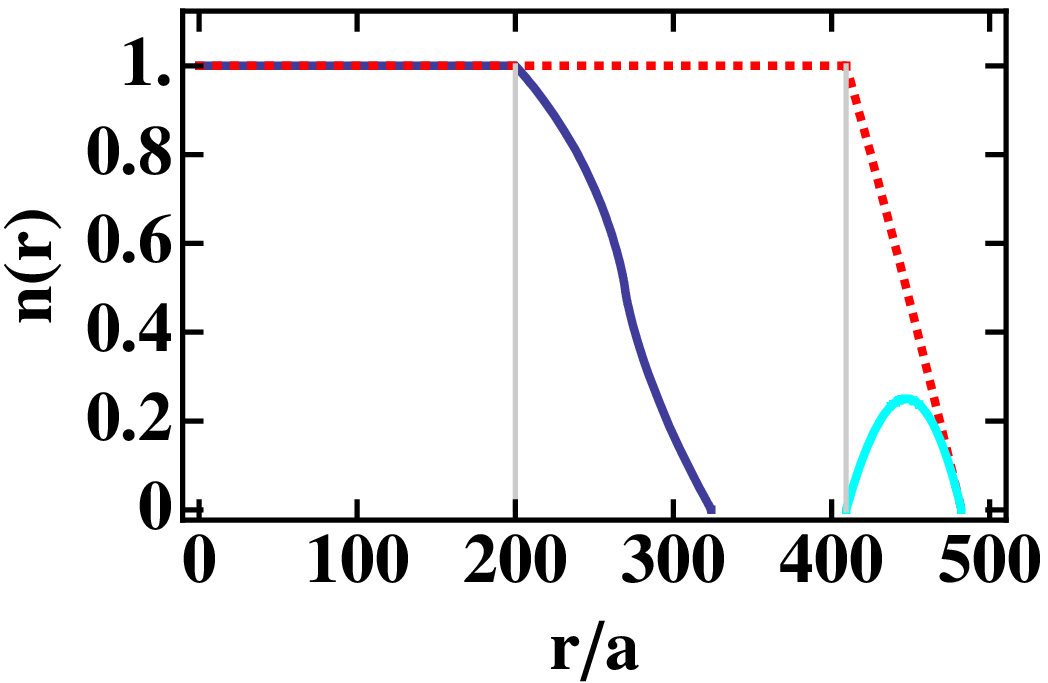}
}  } 
\caption{
\label{fig:bfmix} (color online) 
(a) Shell structure of Bose-Fermi mixtures in harmonically 
confined optical lattices showing a coexisting 
Bose-Mott and Fermi-band insulator region at the center 
$(n_B = 1, n_F = 1)$, a coexisting Bose-Mott insulator and Fermi liquid
region (in blue), a Bose-Mott insulator region $(n_B = 1, n_F = 0)$,
and Bose-Superfluid region at the edge (in red).
(b) filling factors for fermions shown as the solid (dark blue) curve 
and for bosons shown at the dashed (red) curve. 
The solid parabolic curve (light blue) shows 
the order parameter in the superfluid region. 
The parameters used are $t_F = t_B=0.0325U_{BB}$, $U_{BF}=0.1U_{BB}$, 
$\mu_B=0.8U_{BB}$, $\mu_F = 0.4U_{BB}$ and $\Omega_F = \Omega_B=8\times 10^{-6}U_{BB}$,
which are representative of Bose-Fermi mixtures with nearly the same
mass such as $^6$Li and $^7$Li, $^{39}$K and $^{40}$K, or $^{172}$Yb and $^{173}$Yb. 
For the parameters chosen, the widths of the superfluid and FL shells are
several times larger than the lattice spacing $a$.
}
\end{figure}

%
%
The order parameter equation (OPE) is determined 
by minimization of ${\bf E}$ with respect to $\psi^{*}(\mathbf{r})$
leading to 
\begin{equation}
\label{eqn:order-parameter}
\Delta(\mathbf{r})
-\frac{(n_B + 1) t_B (z+a^{2}\nabla^{2})\Delta(\mathbf{r})}
{2\sqrt{\left\vert E_d(\mathbf{r})\right\vert^{2}+
(n_B + 1)\left|\Delta(\mathbf{r})\right|^{2}}}=0.
\end{equation}
Notice that the OPE is not of the Gross-Pitaeviskii (GP) type,
since the superfluid regions emerge from local fluctuations between 
neighboring Mott shells.
Ignoring the spatial derivatives of $\psi$  in Eq.~\ref{eqn:order-parameter} leads to the  
spatially dependent order parameter 
\begin{equation}
\label{eqn:order-parameter-solution}
\left|\psi(\mathbf{r})\right|^{2} = \frac{n_B + 1}{4}
-\frac{\left(n_BU_{BB} + n_F U_{BF} - \mu_B ({\bf r}) \right)^{2}}{4z^{2} t_B^{2}(n_B + 1)}.
\end{equation}
Since $\left|\psi(\mathbf{r})\right|^{2} \ge 0$, hence 
$| n_B U_{BB} + n_F U_{BF} - \mu_{ B,\bf{r} } | \le (n_B +1) z t_B$, and 
the inner $R_{n_B, n_F,-}$ and outer $R_{n_B, n_F, +}$ radii for the superfluid shell 
between the $(n_B, n_F)$ and $(n_B + 1, n_F)$ Mott regions are obtained by setting 
$\left|\psi(\mathbf{r})\right|^{2} = 0$ 
leading to 
%
$$
R_{n_B, n_F, \pm}= R_{B,n_B, n_F} 
\sqrt{ 1 \pm \frac{2 z t_B (n_B + 1)}{\Omega_B} \frac{a^2} { R_{B, n_B, n_F }^2} }.
$$
%
%
%
This relation shows explicitly that $t_B$ splits the spatial degeneracy 
of the $(n_B, n_F)$ and 
$(n_B +1, n_F)$ insulating shells at $r = R_{c,n_B, n_F}$ or 
$\mu_B ({\bf r}) = n_B U_BB + n_F U_{BF}$
by introducing a superfluid region of width
$\Delta R_{n_B, n_F} = R_{n_B, n_F,+} - R_{n_B, n_F,-}$. 
(See Fig.~\ref{fig:bfmix} for characteristic widths).

In addition, the local bosonic filling fraction 
%
$$
n_B (\mathbf{r}) = -\frac{\partial E_-(\mathbf{r})}{\partial
\mu_B }=n_B +\frac{1}{2}-\frac{n_B U_{BB} + n_F U_{BF} - \mu_B ({\bf {r}})} {2z t_B (n_B + 1) }
$$
%
in the same region interpolates between $n_B +1$ for $r \lesssim  R_{n_B, n_F, -}$ and
$n$ for $r \gtrsim  R_{n_B, n_F, +}$, while the chemical potential $\mu_B $ is fixed by 
the total number of bosons $N_B = \int d {\bf r} n ({\bf r})$. 
The local bosonic compressibility 
%
$ 
\kappa_B ( {\bf r} ) = 
\partial n_B ( {\bf r} )/ \partial \mu_B 
= 1 /2 z t_B (n_B + 1)
$
%
of the superfluid shells is non-zero, 
in contrast to the incompressible ($\kappa_B  = 0$) $(n_B, n_F)$ and $(n_B + 1, n_F)$ insulating
shells for  $r < R_{n_B, n_F, -}$ and $r > R_{n_B, n_F, +}$, respectively. 
%

%
%
Now, we consider finite $t_F$. In order to have a tractable theory
we assume that the shell boundaries of the bosons and fermions are well
separated. This allows us to investigate the Fermi liquid near the shell
boundary of the fermions in the presence of a Bose-Mott insulator with
$n_B$ bosons per site. Furthermore, if we assume that the local density
of the Fermi gas is smoothly varying then the local number of fermions is
\begin{equation}
n_F({\bf r}) = \int_{\epsilon_{\rm min}}^{\epsilon_{\rm max}} d \epsilon {\cal D} (\epsilon) f[\epsilon - \mu^{\rm eff}_F({\bf r})] 
\end{equation}
where $f[x]$ is the Fermi function at temperature $T$, and ${\cal
D}(\epsilon)=\sum_{{\bf k}} \delta(\epsilon-\epsilon_{{\bf k}})$ is
the density of fermion states with energy dispersion $\epsilon_{{\bf k}}=-2t_F\sum_\ell^d \cos(k_\ell a)$.  The effective chemical potential
$\mu^{\rm eff}_F({\bf r})=\mu_F({\bf r})-n_B U_{BF}$ accounts for the effect
of the bosons. The band minimum and maximum of $\epsilon_{{\bf k}}$
are $\epsilon_{\rm min}=-2dt_F$ and $\epsilon_{\rm max}=2dt_F$, respectively.
Thus, the Fermi liquid region is limited by the boundaries 
$\epsilon_{\rm min} \le \mu_F ({\bf r}) \le \epsilon_{\rm max}$,
leading to $R_{F, \pm} = R_{F,n_B} \sqrt{ 1 \pm 2 zt_F a^2/\Omega_F R_{F,n_B}^2 }$
for the inner $R_{F, -}$ and outer $R_{F, +}$ radius of the FL shell. 
The width of the FL region is $\Delta R_{F,n_B}  = R_{F,+} - R_{F,-}$.
(See Fig.~\ref{fig:bfmix} for characteristic widths). 
%
%
The isothermal compressibility of the FL region is 
$\kappa_F ({\bf r}) = \partial n_F ({\bf r})/\partial \mu_F$, 
which leads at zero temperature to $\kappa_F ( {\bf r} ) = 0$ 
outside the FL shell, indicating the presence of insulating regions and 
$\kappa_F ( {\bf r} ) = {\cal D} [\mu_F ({\bf r}) ]$ inside the FL shell, 
indicating the presence of conducting regions.
The superfluid and FL shells for finite $t_B$ and $t_F$, 
and their density profiles are shown in Fig.~\ref{fig:bfmix}
for the two-dimensional case.

Next, we propose an experiment to detect superfluid and Fermi liquid shells in 
Bose-Fermi mixtures using a combination of Gaussian and Laguerre-Gaussian beams
followed by Bragg spectroscopy. To illustrate the idea, we discuss the 
simpler case of a nearly two-dimensional configuration, 
where the harmonic trap is very tight along the z-direction, 
loose along the x- and y- directions.
Upon application of Gaussian and Laguerre-Gaussian beams along the z-direction, 
only angular momentum is transferred to the atoms in the 
conducting phases (superfluid or Fermi liquid), imposing a rotating current 
with a well defined velocity profile, while the insulating regions do not absorb angular
momentum due to their large gap in the excitation spectrum.

\begin{figure} [htb]
\centerline{ \scalebox{0.55} {\includegraphics{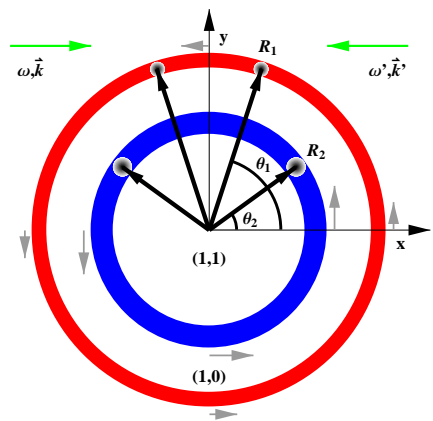}} } 
\caption{\label{fig:brag}
(Color online) Schematic plot for the detection of outer (red) superfluid and 
inner (blue) Fermi liquid shells using Bragg spectroscopy.
The angles $\theta_1$ and $\theta_2$ indicate the locations
of strongest momentum transfer from the Bragg beams (large green arrows) 
to the rotating superfluid and Fermi liquid
shells of radii $R_1$ and $R_2$. The gray arrows indicate 
the sense of rotation of the conducting shells.
}
\end{figure}

To probe the rotating superfluid and Fermi liquid phases we propose the
use of two counter-propagating Bragg beams applied in the xy plane 
along the x direction, as indicated in Fig.~\ref{fig:brag}. 

The Bragg beams transfer a net linear momentum $\hbar(k+k'){\bf x}$ to the
atoms of mass $m$ which satisfy the energy conservation condition
\begin{equation}
\label{eq:bragg}
\hbar(\omega-\omega') = \epsilon_f - \epsilon_i - v_x\hbar(k+k') + \frac{\hbar^2 (k+k')^2}{2m},
\end{equation}
where $v_x$ is the component of the velocity ${\bf v}({\bf r}) = {\bf p}({\bf r})/m$ along the x direction
and $\epsilon_i$ and $\epsilon_f$ are the experimentally accessible 
energies of the initial and final internal states atom.
For an atom carrying one unit of  angular momentum, the velocity is ${\bf v}_i = \hbar\hat{\theta}/mr$. 
Therefore,
within a conducting shell with radius $r=R$ atoms get a linear momentum kick of $\hbar(k+k') \hat{\bf x}$ when the velocity  
$v_x=\hbar\sin\theta/mR$ satisfies the condition given in Eq.~\ref{eq:bragg}. 
This leads to two Bragg angles $\theta=-\sin^{-1}(mRv_x/\hbar)$, and $\pi-\theta$ for each 
conducting shell. As can be seen in Fig.~\ref{fig:brag}, the Bragg angles are $\theta_1$ and $\pi-\theta_1$ 
for the outer superfluid shell
labelled by  $R_1$, and are $\theta_2$ and $\pi-\theta_2$ for the fermi liquid shell
labelled by $R_2$. 
Once these atoms are kicked out of the conducting shells, they form two small expanding
clouds, which can be detected by direct absorption imaging.

Next, we discuss the time scales over which the rotation in 
the conducting regions persist and can be detected experimentally.
In the case of the superfluid region we use the Landau criterion
to show that the velocity imposed to the superfluid through the angular momentum transfer 
is much smaller than the local sound velocity 
$c ({\mathbf r}) = \sqrt{\rho_s ( {\bf r})/ \kappa}$, where 
$\rho_s ( {\bf r}) = 2 t_B a^2 \vert \psi ({\bf r} ) \vert^2$ 
is the local superfluid density, and $\kappa$ is the compressibility.
Thus, $c ({\mathbf r}) = 2 \sqrt{(n_B + 1)z} t a |\psi (\mathbf{r})|$ 
vanishes at the insulator boundaries where $\vert \psi ( {\bf r} ) \vert = 0$, 
and only close to the edge of the superfluid regions the  
local rotational speed $ v ( {\bf r} ) = \hbar / m r $ exceeds $c( {\bf r} )$, which
means that essentially all the superfluid region can be detected and the angular momentum
transferred does not decay over time scales of at least seconds, limited by the lifetime of the trapped system.

In the case of the Fermi liquid region, the time scale over which the flow of the fermions 
persist in presence of the Bose-Mott insulator background can be calculated
from the imaginary part of the fermionic self-energy 
$$
\Sigma_F (k)  =  U_{BF}^2 T^2\sum_{q_1, q_2} G_B (q_1)
G_B (q_2) G_F (k +q_1 - q_2),
$$
where $k = ({\bf k}, i\omega)$ and $q_i = ({\bf q}_i, i\nu_i)$, with 
$\omega$ $(\nu_i)$ are fermionic (bosonic) Matsubara frequencies and $T$ is temperature.
The bare inverse bosonic propagator in the Bose-Mott phase is 
%
$$
G_B^{-1} (q, {\bf  r}) = 
\epsilon_{{\bf q}} 
\left[ 1 + 
\epsilon_{\bf q} \left( \frac{n_B +1}
{i\hbar\omega - E_1 ({\bf r})}  - \frac{n_B}{i \hbar\omega - E_2 ({\bf r})} \right) \right],
$$
%
%
where $E_1({\bf r}) = (n_B-1)U_{BB} - \mu_B({\bf r})$,
$E_2({\bf r}) = (n_B-2)U_{BB} - \mu_B({\bf r})$ and
$\epsilon_{\bf q} = - 2 t_B \sum_{\ell}^{d} \cos ( k_{\ell} a )$.
The bare inverse fermionic propagator in the Fermi liquid phase 
is 
$
G_F^{-1} (k, {\bf r}) = 
i \hbar \omega - \epsilon_F ({\bf k}, {\bf r}), 
$
where $\epsilon_F ({\bf k}, {\bf r}) = \epsilon_F ({\bf k}) - \mu_F ({\bf r})$.
For $n_B = 1$ and $T=0$, the imaginary part of the fermionic self-energy is 
\begin{equation}
\label{eqn:self-energy}
{\rm Im} \Sigma_F (k, {\bf r}) 
= -\pi {U^2_{BF}} \left[ F (\hbar\omega) + F (-\hbar\omega) \right],
\end{equation}
where $F (\hbar \omega) = 
\Theta (\hbar\omega) \Theta(-\hbar\omega + \mu_B ({\bf r}))
{\cal D} ( \hbar\omega + \mu_F ({\bf r}))$, $\Theta$ is
the Heaviside step function, and 
${\cal D} (\epsilon)$ is the Fermion density of 
states. For a two-dimensional Fermi liquid shell, there is a Van Hove singularity 
in ${\cal D} (\epsilon)$ at half filling.
The expression in Eq.~(\ref{eqn:self-energy}) is independent of momentum, 
since the dominant excitations in the Bose-Mott region 
are number-conserving and low-momentum particle-hole excitations, 
but strongly dependent on position through $\mu_B ( {\bf r} )$ and 
$\mu_F ( {\bf r} )$, leading to a characteristic decay time
$\tau_{\bf r} = - h/{\rm Im} \Sigma (k, {\bf r})$.
For the parameters used in Fig.~\ref{fig:bfmix},
(with $U_{BF}/h = 1~{\rm kHz}$) the time scale for the persistence of 
the flow near the edges (away from the Van Hove singularity) is 
$\tau_{\bf r} \approx 13~{\rm ms}$. However, near the center of the 
Fermi liquid region (close to the Van Hove singularity) $\tau_{\bf r}$ is 
extremely short, indicating that it is much easier to detect fermions 
at the edge than at the center of Fermi liquid shells.

%
%

We have discussed the phase diagram of Bose-Fermi mixtures in harmonically
confined optical lattices in the regime where the hopping parameters of bosons
and fermions are comparable and the repulsion between bosons and fermions is a
substantial fraction of the boson-boson repulsion. We showed that the system
exhibits regions of (I) coexisting Bose-Mott and Fermi-band insulator, 
(II) coexisting Bose-Mott insulator and Fermi liquid, (III) Bose-Mott insulator,
and (IV) Bose-superfluid. We have calculated analytically 
the boundaries between these phases and obtained the spatially dependent filling
fraction for each region. Finally, we proposed a detection method of the 
superfluid and Fermi liquid shells of Bose-Fermi mixtures by using Gaussian and
Laguerre-Gaussian beams followed by Bragg spectroscopy.

\end{document}